\documentclass[preprint,aps]{revtex4}
\usepackage{graphicx}
\begin{document}
\title{Can extra dimensional effects replace dark matter ?}
\author{Supratik Pal \footnote{Electronic address : {\em supratik@cts.iitkgp.ernet.in};
Permanent address : Department of Physics, 
Ramakrishna Mission Vidyapith, Purulia 723 147 
and Department of Physics, Jadavpur University, Kolkata 700 032, India},
${}^{}$ Somnath Bharadwaj \footnote{Electronic address : {\em somnath@cts.iitkgp.ernet.in}}
${}^{}$
and Sayan Kar \footnote{Electronic address : {\em sayan@cts.iitkgp.ernet.in}}
${}^{}$}
\affiliation{Department of Physics and Centre for
Theoretical Studies\\ Indian Institute of Technology\\ Kharagpur 721 302, India}
\vspace{.5in}

\begin{abstract}
In the braneworld scenario,  the four dimensional   effective Einstein
 equation  has extra terms which arise from the  embedding of the
 3-brane  in the bulk.  We show that in this  modified theory
 of gravity, it is possible to  model
 observations of 
 galaxy rotation curves and the X-ray profiles of clusters of
 galaxies, without the need for dark matter. In this scenario, a  
 traceless tensor field which arises
 from the projection  of the bulk Weyl tensor on the brane, provides
 the extra gravitational  acceleration which is usually explained
 through dark matter. We also predict that gravitational lensing
 observations can possibly discriminate between the proposed higher
 dimensional  effects and dark matter, the deflection angles predicted
 in the  proposed  scenario  being around  $75 \%$ to $80 \%$ of the
 usual  predictions  based on  dark matter.        
\end{abstract}

\maketitle

\noindent {\em Introduction :} 
To determine the nature of dark matter {\cite{dark}} and how it is distributed
is one of the most important problems  currently facing physicists and 
astro-physicists. The  problem arises, over a range of astrophysical
length-scales,  from a  variety of  observations which determine the 
dynamical  mass. Observations of galaxy rotation curves {\cite{curv}} 
and the gravitational lensing {\cite{sef}}
by galaxies are some of the most direct probes of dark matter on
galactic scales ($\sim 10 - 100\,  {\rm kpc}$).  The X-ray profiles of
clusters of galaxies and the gravitational lensing by these
objects  probe dark matter on   larger scales $(\sim 0.1 - 
10 \, {\rm Mpc})$ {\cite{clusters}}. 

The usual analysis of these observations is based on two assumptions :
\begin{itemize}
\item[1.] Einstein's theory of gravitation as given by the equations
  $G_{\mu \nu}=(8  \pi G/c^4) T_{\mu \nu}$ is valid on the
  length-scales in question. 
\item[2.] The matter in galaxies and clusters of galaxies is such that
  relativistic stresses do not 
make a significant   contribution to the stress-energy tensor $T_{\mu
  \nu}$, and $T_{00}=\rho c^2$ is the only non-zero component. 
\end{itemize}

It follows that the Einstein tensor $G_{\mu \nu}$ has only one
non-zero component $G_{00}$ which can be determined directly from 
observations of  either the rotation curve or the X-ray profiles,  or
from  gravitational lensing, and  this allows the total matter
distribution to be mapped  out. The mass determined from such
dynamical means  is {\it always} found to be  in excess of that which
can be attributed to the visible matter. This discrepancy is explained
by postulating that every galaxy and cluster of galaxy is embedded in
a halo made up of some kind of invisible matter, the dark matter {\cite {dark, 
curv, wimp}}. 
The exact nature of this dark matter is unknown, with exotic
supersymmetric particles {\cite{wimp}} currently being accepted as the most
favoured candidates. Till date, direct searches for the dark matter 
particles have not yielded any detection.

  It is important to take note of the fact that none of the
  assumptions used  have been independently tested on either the
  galactic or the cluster length-scales. This raises the possibility
  that there actually may not be any dark matter and it may be
  possible to explain these observations using a modified theory of
  gravity {\cite {mond}}. Another possibility is that the dark matter may have
  relativisitic stresses, which will require a different interpretation
  of the observations and will result in a different inferred dark
  matter distribution {\cite {lake, bk}}. 

The possibility that we live in a warped  five dimensional (or
possibly  higher) space-time, in which  the familiar 4-dimensional
space-time is a   hyper-surface {\cite {rs}} has, of late,  received a
considerable amount of attention {\cite{csaki}}. In this  so-called {\it
  brane-world}  scenario, the effective Einstein equations pick up
extra terms {\cite {shiro}} which arise from the embedding of our four dimensional
space-time hyper-surface, referred to as the   
brane, in the five dimensional space-time, the bulk. In this {\it
  Letter}, we show 
that it is possible to model the halos of galaxies and clusters of
galaxies using the modified Einstein equation, and thereby explain
the observed rotation curves and X-ray profiles without the need for
dark matter.    

The effective four dimensional Einstein equations {\cite {shiro}}
is given by : 

\begin{equation}
G_{\mu \nu} = -\Lambda g_{\mu \nu} + \kappa_4^2 T_{\mu \nu} +
\kappa_5^4 S_{\mu \nu} -E_{\mu \nu}
\label{eq:a1}
\end{equation}

where $\Lambda$,  the brane cosmological constant, depends on the bulk
cosmological constant and the brane tension $\lambda_b$ both of which
can be fine-tuned to make $\Lambda=0$ which we adopt throughout. The
constants $\kappa_4$ and $\kappa_5$ are defined as $\kappa_4^2=8 \pi
G/ c^4=  \kappa_5^2 \lambda_b /6$,  and  $E_{\mu \nu}$ is the limit,
of the projection on the brane, of a  quantity defined in 5-dimensions
{\cite{shiro,pdl}},   which is related to the bulk Weyl tensor and the
bulk matter.    In the absence of bulk matter,  $E_{\mu\nu}$ is a
traceless symmetric tensor {\cite {mrt, dad, mak}}. 
The term $S_{\mu \nu}$ is   quadratic in the brane energy-momentum
tensor, and  it can be shown to be small compared to both the
usual linear energy-momentum tensor $T_{\mu \nu}$  and  $E_{\mu\nu}$
{\cite{shiro}}.   In the following, we shall ignore the contribution
from  the term $S_{\mu \nu}$. 

To summarize, the effective Einstein equation on the brane is 
\begin{equation}
G_{\mu \nu} =-E_{\mu \nu} +  \kappa_{4}^2 T_{\mu\nu}
\label{eq:a2}
\end{equation}
where the difference from the usual Einstein equation is that we have
an extra traceless tensor $E_{\mu \nu}$ which is a purely geometrical
term that arises from  embedding of the 3-brane in the bulk. We
investigate if it is possible to consistently model observations of
galaxies and clusters of galaxies without the need for dark matter  in
this modified theory of gravity.

{\em Modelling galaxy and cluster halos:}
It is possible to interpret observations of the rotation curves of
spiral galaxies and the X-ray profiles of clusters of galaxies without
reference to any particular theory of gravity or the existence and
nature of dark matter, the only assumption being that gravitation
arises from the geometry of space time and the gravitational field can
be represented by the space-time metric $g_{\mu \nu}$ which we choose
with signature $(-,+,+,+)$. In the work
presented here we make two further assumptions which, though not
crucial for the discussion, substantially simplify the analysis.  

First, we assume that galaxies and clusters of galaxies are embedded
inside spherically symmetric gravitational fields which we refer to as 
halos. The most general static, spherically symmetric space-time is
completely described  by only two unknown  functions 
$\Phi(r)$ and $\Psi(r)$ which we choose so that the proper time inside
the halo is  
\begin{equation}
c^2 d \tau^2 = - (1 + 2 \Phi) c^2 d t^2 + (1- 2 \Phi + 2 \Psi) [d r^2
  + r^2 ( d  
  \theta^2 + \sin^2 \theta \, d \phi^2 )]
\label{eq:a3}
\end{equation}

Further, the gravitational field is assumed to be weak $(\Phi, \Psi
\ll 1)$, and we 
retain terms only  to  linear order in the potentials $\Phi$ and
$\Psi$.  This is an assumption which we shall justify later.  

We next briefly discuss how observations of the rotation curves of
spiral galaxies and the X-ray profiles of clusters of galaxies can be
used to determine the gravitational potentials  inside the halo. 

Observations of the 21 cm line from neutral hydrogen (HI) clouds in
spiral galaxies show these clouds to be distributed in a disk,
aligned with the plane of the galaxy. 
 Further, the observed redshifts of the 21 cm emission 
determine the velocities of the HI clouds. These observations  show
the clouds to be  in circular orbits around the center of the galaxy.
The circular velocity $v_c(r)$ of the HI clouds at
different radius  $r$ from the center is referred to as the ``rotation
curve''.  
The geodesic equation for the HI clouds, which we treat  as test
particles moving in stable circular orbits under the gravitational
influence of the halo,    
\begin{equation}
\Phi^{'}(r)=\frac{1}{r} \frac{v^2_c(r)}{c^2}
\label{eq:a4}
\end{equation}
determines the potential $\Phi(r)$ in terms of the observed rotation
curve. The rotational velocities are typically in the range $(100-400)
{\rm km/s}$ and as a consequence $\Phi \sim (v_c/c)^2 \sim  10^{-6}$,
validating our initial assumption  that the gravitational field is
weak.  

Interpreting the observed X-ray emission from the hot, ionized
intra-cluster gas in clusters of galaxies under the assumption that
the gas is isothermal  
allows the density profile $\rho_g(r)$, and the pressure $P_g(r)= (k
T/\mu m_p)\,  \rho_g(r)$ of the gas to be determined. Here $k$, $T$, $\mu$
and $m_p$ are the Boltzmann constant, gas temperature, mean atomic
weight of the particles in the gas and the proton mass respectively.
The gas temperatures are typically in the range $2 \times 10^7 - 10^8
\, {\rm K}$ which implies that $(k T/\mu m_p
c^2) \sim 10^{-5} - 10^{-6}$. Considering the energy-momentum tensor
for the intra-cluster gas $T[{\rm gas}]_{\mu \nu}=(P_g + \rho_g c)
U_{\mu} U_{\nu} - P_g g_{\mu \nu}$, we see that $P_g = (k T/\mu m_p)
\rho_g \ll \rho_g c^2$.  Assuming  the gas to be in  hydrostatic
equilibrium,  the energy momentum  conservation 
${T[{\rm  gas}]^{\mu}}_{\nu;\mu} =0$ gives 
\begin{equation}
\Phi^{'}(r)=-\frac{ kT}{\mu m_p c^2}  \frac{ d \ln \rho_g}{d
  r} \,.
\label{eq:a5} 
\end{equation}
which can be used to determine $\Phi$. It should be noted that in the
energy-momentum conservation equation we have dropped terms of order
$P_g \Phi$ and $P_g \Psi$, as these are much 
smaller compared to the terms involvong $\Phi \rho_g c^2$ and $P_g$  which
we have retained.  The factor $(k T/\mu m_p c^2)$ in equation
(\ref{eq:a5}) ensures that $\Phi \ll 1$, justifying the assumption
that the field is weak. 

While it requires two potentials $\Phi$ and $\Psi$ to completely
specify the gravitational field inside the halo,  only one
of the potentials, namely $\Phi$, can be determined directly from
observations of either the rotation curve or the X-ray profiles.  It
should 
also be noted that $\Phi$ is the only one which matters if we are
dealing with the motion of non-relativistic $(v/c \ll 1)$ particles. 
The potential $\Psi$ is important when considering  the motion of
relativisitic particles, eg. photons which  we shall consider later.   

It is necessary to assume a specific  theory for gravity if one is to
proceed further in modelling galaxy or cluster halos. We shall first
briefly outline the standard procedure which is based on Newtonian
physics {\it ie.} the two assumptions mentioned in the Introduction.   
The components of the Einstein tensor (which appear 
in nearly all geometrical theories for gravity) are listed below.
\begin{equation}
G^0_0=- 2 \nabla^2 (\Phi - \Psi) \,, \,  G^r_r=2 \frac{\Psi^{'}}{r} \,
, \, 
G^{\theta}_{\theta}=G^{\phi}_{\phi}=\Psi^{''}+\frac{\Psi^{'}}{r}
\label{eq:a10}
\end{equation}
 Under the abovestated
Newtonian assumptions {\it ie.} Einstein's theory is valid
on these length-scales, we have
\begin{equation}
G^{\mu}_{\nu}=\frac{8 \pi G}{c^4} T^{\mu}_{\nu}
\label{eq:a11}
\end{equation} 
and that relativisitc stresses are absent in $T^{\mu}_{\nu}$ and
$T^0_0=-\rho c^2$ is the only non-zero component, implies $\Psi=0$ and
\begin{equation}
\nabla^2 \Phi = \frac{4 \pi G}{c^2} \rho
\label{eq:a12}
\end{equation} 
 
The reader has probably already realised that $ c^2 \Phi$ is
the familiar  gravitational potential which appears in Newtonian
gravity,  and $\Psi$ quantifies deviations from the Newtonian theory.  

In the Newtonian theory, the dark matter problem arises when one uses
the potential $\Phi$ determined from observations of either the
rotation  curves (eq. \ref{eq:a4}) or the  X-profiles
(eq. \ref{eq:a5}) in equation  (\ref{eq:a12}) to make estimates of
the density.   These dynamical estimates of the  density and the mass
are always found to be substantially in excess of the visible matter.  
Hence it is required to  postulate that around $\sim 80 \%$, or more,
of the matter in the outer parts of   spiral galaxies and in clusters
of galaxies is invisible, {\it ie.}  the dark matter.   

We next consider the  modified theory of gravity as discussed in
equation (\ref{eq:a2}), where the Einstein
equation has an extra traceless term $E^{\mu}_{\nu}$ arising from
the embedding of the 3-brane in higher dimensions. 
\begin{equation}
G^{\mu}_{\nu}+E^{\mu}_{\nu}=\frac{8 \pi G}{c^4} T^{\mu}_{\nu} \,. 
\label{eq:a13}
\end{equation} 
We proceed by taking the trace of eq. (\ref{eq:a13}) which gives us 
\begin{equation}
\nabla^2 (\Phi - 2 \Psi)= \frac{4 \pi G}{c^2 }\rho_v
\label{eq:a14}
\end{equation}
where we have assumed that there is no dark matter,  and the visible
matter with density $\rho_v$ is all that contributes to the
energy-momentum tensor. The solution to equation (\ref{eq:a14}) is  
\begin{equation}
\Psi=\frac{1}{2} \Phi - \frac{2 \pi G}{c^2} (\nabla^2)^{-1} \rho_v \, 
\label{eq:a15}
\end{equation}
where $\Phi$ is determined from  observations of either rotation
curves or 
the X-ray profiles.  The tensor $E^{\mu}_{\nu}$ can be calculated
using eq. (\ref{eq:a13}) once both $\Phi$ and $\Psi$ are known. 

 The point to note here is that we have a solution for the gravitational
 field inside the halo, consistent with observations of rotation
 curves or X-ray profiles, without the need for dark matter. The extra
 gravitational acceleration  required to explain observation of galaxy 
 rotation curves or the cluster X-ray profiles  now arises from
 $E^{\mu}_{\nu}$ which   incorporates the geometrical effects arising
 from the embedding of  the 3-brane  in the bulk. The proposal that
 $E^{\mu}_{\nu}$ can replace  dark matter has
 been made earlier  \cite{mak},  but not  substantiated in a general
 situation. The earlier work imposes an ad hoc conformal symmetry to
 obtain spherically symmetric vacuum solutions of the modified
 Einstein's equations (eq. \ref{eq:a2}) which are consistent
 with flat rotation curves. In this {\it Letter} we have  outlined, 
 in general and without any ad hoc assumptions,   how rotations curves
 and X-ray profiles can be interpreted  without  dark matter 
in the modified theory of gravity.   

We next take up a specific example and explicitly calculate 
$\Phi$ and $\Psi$. Interpreting observations of  spiral galaxies is
somewhat complicated as the visible matter is mainly distributed in 
a disk \cite{gal1}  which breaks spherical symmetry.  Further,
ambiguities in the mass to light ratio \cite{mtol} makes it difficult
to uniquely  determine the mass corresponding to  the  visible
matter. The situation is simpler for  clusters where the X-ray gas is
the dominant component of visible matter ({\it ie}
$\rho_v=\rho_g$). The  mass density of X-ray gas is usually modeled  
using the spherically symmetric, isothermal $\beta$ model with
\begin{equation}
\rho_g(r)=\rho_0 [1 + (r/r_c)^2]^{-3 \beta/2} 
\end{equation}
where $\rho_0$ is the central density, $r_c$ the core radius and
$\beta$ decides the slope at $r \gg r_c$, and these parameters  have
values in the 
range $(7 - 150)\,  \times 10^{-23} {\rm kg/m^{3}}$, $0.1 - 0.8 \,
{\rm Mpc}$ and $0.5-0.9$ respectively \cite{clusterd}. 
For simplicity we use $\beta=2/3$ and  restrict  our analysis to $r
\gg r_c$. Further, we use $\rho_0=5 \times 10^{-24} {\rm kg/m^3}$,
$r_c=0.3 \, {\rm Mpc}$, $\mu=0.6$ and $T=10^8 {\rm K}$ as
represntative values when making estimates. 

 Solving equation (\ref{eq:a5})  gives us 
\begin{equation}
\Phi = \frac{2kT}{\mu m_p c^2} \ln \frac{r}{r_c} \,.
\label{eq:c1}
\end{equation}
In the usual Newtonian analysis $\Psi=0$, and $\Phi$ 
 is used  in  eq. (\ref{eq:a12})  to determine the total matter
 density  $\rho(r)=(k   T/2 \pi G \mu m_p) \,  r^{-2}$  needed to keep
 the hot X-ray gas in hydrostatic  equilibrium.  Comparing $\rho(r)$
 with  $\rho_g(r)$  we find that 
 $\rho_g(r)/\rho(r)=(2 \pi  G \rho_0 r_c^2 \mu m_p/k T) \sim 0.2$  {\it
 ie.} $80 \%$ of the total matter has to be in an invisible form, the
 dark matter. 

In the modified theory $\Psi \neq 0$ and we solve eq. (\ref{eq:a15}) to
obtain  
\begin{equation}
\Psi = \left [ \frac{kT}{\mu m_p c^2}-\frac{2\pi G\rho_o r_c^2}{c^2} \right ]
\ln \frac{r}{r_c} 
\label{eq:c2}
\end{equation}
which is a solution of the modified Eunstein's equations
(eq. \ref{eq:a13}) without any dark matter. The non-zero
components of  $E^{\mu}_{\nu}$,  $E^0_0=-E^r_r=[(2
  kT/\mu m_p c^2)-(4 \pi G \rho_0 r_c^2/c^2)] r^{-2}$ now  
   provide the extra gravitational acceleration. 

The  two different theories for gravity considered here interpret 
the same X-ray observations to infer different space-time geometries
for cluster  halos. It is necessary to consider other  independent
probes of the space-time geometry to  discriminate between the two
possibilities namely dark matter and higher dimensional 
effects.  

\noindent {\em Gravitational Lensing:} 
Observations of gravitational lensing {\cite{sef}}
provide independent constraints on the gravitational field inside 
halos. These observations probe both $\Phi$ and $\Psi$, and are
sensitive to the full geometry of the  space-time inside halos. 
  In the standard Newtonian analysis where
  $\Psi=0$, the deflection angle $\hat{\alpha}_N$ of a photon 
from a distant  source ($s$), propagating through the halo  to a distant
observer ($o$) is given to be 
\begin{equation}
\hat{\alpha}_N = 2 \int_{s}^{o}  \hat{\nabla}_\perp  \Phi
\, \,  dl \, 
\label{eq:b1}
\end{equation}
where the integral is to be evaluated along the straight line
trajectory between the source and the observer, and
$\hat{\nabla}_\perp$ denotes 
the derivative in the direction perpendicular to this
trajectory. Using eq. (\ref{eq:c1}) we find that a photon passing
through the halo of a cluster experiences a 
constant deflection  given by 
\begin{equation}
\alpha_N=\frac{4\pi kT}{\mu m_p c^2}
\label{eq:b2}
\end{equation}
Generalizing eq. (\ref{eq:b1}) to the situation where $\Psi 
\neq 0$ gives  
\begin{equation}
\hat{\alpha } =  \int_{s}^{o} \hat{\nabla}_\perp (2 \Phi -
\Psi)  \, \,  dl  \,.
\label{eq:b3}
\end{equation}

Using  $\Phi$ and $\Psi$ calculated for the modified theory of
gravity (eqs. \ref{eq:c1} and \ref{eq:c2}) gives the deflection angle
to be 
\begin{equation}
\hat{\alpha } = .\hat{\alpha_N} \left[ 0.75 + \frac{\pi G \rho_0
    r_c^2 \mu m_p}{2 k T}\right]  
\label{eq:b4}
\end{equation}

The term $\frac{\pi G \rho_0 r_c^2 \mu m_p}{2 k T}$ which arises from
the contribution of the visible matter to $\Psi$ (eq. \ref{eq:a15}) is
around $0.05$ for our choice of cluster parameter and it falls to less
than $0.01$ if $r_c=0.1 \, {\rm Mpc}$.

We find that  the modified theory with no dark matter predicts a 
lensing deflection angle which is  smaller  than  that of the usual 
Newtonian  analysis where there is dark matter. For the cluster
parameters adopted here, it is  $80 \%$ of the Newtonian value, and it
is expected to be in the range $75 \%$ to $80 \%$ of the Newtonian
value for  typical clusters, depending on the cluster parameters.
This should, in principle, allow us to observationally  
discriminate between the two possibilities and determine which is
correct. Carrying this out requires  X-ray and gravitational lensing
observations of the same cluster. The X-ray profiles can be used to 
determine the metric which will be different in the two
scenarios. These can be used  to make lensing predictions which can be 
compared with observations to test which scenario is correct.  There
presently exists a substantial volume of  such observations which have
been interpreted in the Newtonian picture using dark matter. 
A significant fraction of these observations have
been interpreted to conclude that the  dark matter masses inferred
from X-ray observations  are significantly smaller ($\sim$ 2 to 4
times) than the masses inferred from gravitational lensing
\cite{clusterlens1}, while  there also are a   significant number of
claims that  the X-ray and lensing observations  are consistent
\cite{clusterlens2}. At a preliminary level it may be speculated that  
the present uncertainities  (statistical and  systematic) in the
modeling of X-ray and lensing  observations are sufficiently large
that  the  alternate possibility considered here, whose predictions
differ by around $20   \%$ ,  would fare equally well as  the dark
matter   scenario in  simultaneously fitting  X-ray and lensing
data. 

There currently exists a large body of observations on cosmological
scales (1 Mpc to 10 Gpc) like the CMBR anisotropies {\cite{cmbr}} and
the clustering of galaxies \cite{sdss}  all of which are 
consistent, at a high level of precision, with a cosmological model
where one third of the present matter density is in cold dark matter
and two-thirds in dark energy which has negative pressure, the sum of
the two densities being very 
close to the critical value $3 H^2_0/8 \pi G$,  where $H_0$ is the
present value of the Hubble parameter. The interpretation of these
observations requires the analysis of the growth of perturbations in
an expanding background cosmological model. It is to be seen if the
dynamics of perturbations in the presence of $E_{\mu \nu}$ can explain
these observations without the need for dark matter.

\end{document}